\documentclass[aps,prb,twocolumn,superscriptaddress,showpacs,showkeys]{revtex4-2}

\usepackage{xr}
\usepackage{graphicx}
\usepackage{color}
\usepackage{epstopdf}
\usepackage{amsmath}
\usepackage{braket}
\usepackage[colorlinks=true, linkcolor=blue, urlcolor=blue, citecolor=blue]{hyperref}

\usepackage{changes}

\begin{document}

\newcommand{\ALBA}[0]{{
ALBA Synchrotron Light Source, carrer de la Llum 2-26, 08290 Cerdanyola del Vall{\`e}s, Spain}}
\newcommand{\CFM}[0]{{
Centro de F\'{\i}sica de Materiales CSIC-UPV/EHU-Materials Physics Center, 20018 San Sebasti\'an, Spain}}
\newcommand{\UPVPMA}[0]{{
Departamento de Pol\'imeros y Materiales Avanzados: F\'isica, Qu\'imica y Tecnolog\'ia,
Universidad del Pa\'is Vasco UPV/EHU, 20018 San Sebasti\'an, Spain}}
\newcommand{\DIPC}[0]{{
Donostia International Physics Center, 20018 Donostia-San Sebasti\'an, Spain}}
\newcommand{\UPVFA}[0]{{
Departamento de F\'isica Aplicada I,
Universidad del Pa\'is Vasco UPV/EHU,  20018 San Sebasti\'an, Spain}}
\newcommand{\IOMCNR}[0]{{
IOM-CNR, Laboratorio TASC, Strada Statale 14 Km 163.5, I-34149 Trieste, Italy}}
\newcommand{\Elettra}[0]{{
Elettra - Sincrotrone Trieste S.C.p.A., Strada Statale 14 Km 163.5, I-34149 Trieste, Italy}}
\newcommand{\TUDD}[0]{{
Institut f\"ur Festk\"orper- und Materialphysik, Technische Universit\"at
Dresden, D-01062 Dresden, Germany}}
\newcommand{\ICN}[0]{{
Catalan Institute of Nanoscience and Nanotechnology (ICN2), CSIC and BIST, Campus UAB, 08193 Barcelona, Spain}}
\newcommand{\ICREA}[0]{{
ICREA-Instituci\'o Catalana de Recerca i Estudis Avan\c{c}ats, Lluis Companys 23, 08010 Barcelona, Spain}}
\newcommand{\IMDEA}[0]{{
Instituto Madrile\~no de Estudios Avanzados, IMDEA Nanociencia, Calle Faraday 9, 28049 Madrid, Spain}}
\newcommand{\Chalmers}[0]{{
Chalmers University of Technology, G{\"o}teborg, Chalmersplatsen 4, 412 96 G{\"o}teborg, Sweden}}
\newcommand{\Nanogune}[0]{{
CIC nanoGUNE-BRTA, 20018 San Sebasti\'an, Spain}}

\author{Mar\'{\i}a Blanco-Rey}
\affiliation{\UPVPMA}
\affiliation{\DIPC}
\affiliation{\CFM}
\author{Rodrigo Castrillo}
\affiliation{\CFM}
\author{Khadiza Ali}
\affiliation{\DIPC}
\affiliation{\Chalmers}
\author{Pierluigi Gargiani}
\affiliation{\ALBA}
\author{Max Ilyn}
\affiliation{\CFM}
\author{Michele Gastaldo}
\affiliation{\ICN}
\affiliation{J. Heyrovsk{\'y} Institute of Physical Chemistry, Czech Academy of Sciences, Dolej{\v s}kova 2155/3, Prague, 18223, Czech Republic }
\author{Markos Paradinas}
\affiliation{\ICN}
\affiliation{Institut de Ci{\`e}ncia de Materials de Barcelona, ICMAB-CSIC, Campus UAB, 08193 Bellaterra, Spain}
\author{Miguel A. Valbuena}
\affiliation{\ICN}
\affiliation{\IMDEA}
\author{Aitor Mugarza}
\affiliation{\ICN}
\affiliation{\ICREA}
\author{J. Enrique Ortega}
\affiliation{\UPVFA}
\affiliation{\CFM}
\affiliation{\DIPC}
\author{Frederik Schiller}
\affiliation{\CFM}
\affiliation{\DIPC}
\author{Laura Fern\'andez}
\affiliation{\Nanogune}
\affiliation{\CFM}
\email{lauraisabel.fernandez@ehu.es}

\title{The role of rare-earth atoms in the anisotropy and antiferromagnetic exchange coupling at a hybrid metal-organic interface}

\pagebreak

\date{\today}

\begin{abstract}
Magnetic anisotropy and magnetic exchange interactions are crucial parameters that characterize the hybrid metal-organic interface, key component of an organic spintronic device. We show that the incorporation of 4$f$ RE atoms to hybrid metal-organic interfaces of CuPc/REAu$_2$ type  (RE= Gd, Ho) constitutes a feasible approach towards on-demand magnetic properties and functionalities. The GdAu$_2$ and HoAu$_2$ substrates differ in their magnetic anisotropy behavior. Remarkably, the HoAu$_2$ surface boosts the inherent out-of-plane anisotropy of CuPc, owing to the match between the anisotropy axis of substrate and molecule. Furthermore, the presence of RE atoms leads to a spontaneous antiferromagnetic (AFM) exchange coupling at the interface, induced by the 3$d$-4$f$ superexchange interaction between the unpaired 3$d$ electron of CuPc and the 4$f$ electrons of the RE atoms. We show that 4$f$ RE atoms with unquenched quantum orbital momentum ($L$), as it is the case of Ho, induce an anisotropic interfacial exchange coupling.
\end{abstract}

\newpage
\maketitle

\section{Introduction}
Molecular spintronics is an emerging field that combines ferromagnetic materials with organic or metal-organic semiconductors.
It benefits from the unique and exceptional properties of organic molecules, which go beyond inorganics~\cite{Cornia2017,Cinchetti2017_NatMat}.
Furthermore, the strong response of many organic molecules to electrical, optical or magnetic stimuli bring new potential functionalities to the
spintronic device~\cite{Comstock2007_PRL,Neguyen2012_Science,Sun2016_Science}. An effective spin tunneling between the ferromagnet and the organic
molecule is governed by the so-called \emph{spinterface} that is formed by the hybrid interface between both materials. The spinterface is defined on the
one hand by the energy level alignment between molecule and ferromagnet, and on the other hand by its anisotropy and the interfacial magnetic
exchange interaction~\cite{Urdampilleta2011NatMat,Neguyen2012_Science,Lach2012,Djeghlouli13,AlMaMari15,Barraud10,Xiong}.
It is thus desirable to be able of controlling both energy level alignment and magnetic interaction. The former can be tuned by choosing the molecular species having convenient functional groups. Tailoring magnetic interactions is more difficult. Several authors have modified the interfacial exchange interaction between a ferromagnetic electrode and organic molecules by the incorporation of spacers like graphene, oxygen atoms or a Cu layer in systems formed by phtalocyanine or porphyrin molecules adsorbed on ferromagnetic Ni or Co surfaces
~\cite{Avvisati2018,Bernien2009, Gruber2015, bib:halder23}. Thus, the natural ferromagnetic (FM) interfacial exchange interaction has been turned to an antiferromagnetic (AFM) one, which is a relevant feature in the design of molecular spintronic devices~\cite{Hyuk-JaeJang}. Another strategy to get an interfacial AFM exchange coupling is to use 3$d$ transition metal surfaces and RE-centered metal-organic molecules~\cite{bib:lodi11,Lodi2014_SS}.

In this work, we explore an alternative way of achieving an interfacial AFM exchange interaction combining a ferromagnetic single atom thick layer that incorporates RE atoms with a single monolayer of organic molecules that include transition metal atoms.  For this purpose, we adsorb copper phthalocyanine (CuPc) molecules on ferromagnetic GdAu$_2$ and HoAu$_2$ single atomic layers~\cite{Ormaza_PRB2013,Corso2010_PRL,Ormaza_NanoLett2016,Fernandez20Nanoscale,BlancoRey2022_PRR}. A spontaneous antiparallel alignment between the CuPc spin and the magnetization of the REAu$_2$ surfaces is detected, which is induced by a superexchange interaction between the unpaired 3$d$ electron of CuPc and the 4$f$ electrons of the RE. Such interfacial AFM coupling has been already observed in different systems like Co nanodots  grown on GdAu$_2$ MLs~\cite{Fernandez2014_NL,Cavallin2014,fernandez2010,Fernandez16_AdvScience} or RE adatoms on Fe islands~\cite{Coffey}. Campbell proposed a model~\cite{Campbell1972} to explain the 4$f$-3$d$ exchange found in RE-transition metal compounds, extensively studied in the past. This model featured an intra-atomic FM exchange interaction between the  RE 4$f$ and RE 5$d$ spins and a subsequent AFM exchange coupling between the itinerant RE 5$d$ and the 3$d$ spins of the transition metal. Furthermore, we have observed in our system that the presence of 4$f$ electrons with an unquenched quantum orbital momentum $L$, as it is the case of Ho, introduces strong spin-orbit effects that lead to an anisotropic exchange mechanism. This unconventional feature is of interest for the development of interfaces with improved functionalities, such as an anisotropic magnetoresistance behavior~\cite{Barraud_2015}.

Both substrate formed by GdAu$_2$ and HoAu$_2$ monolayers (MLs) exhibit a RKKY exchange coupling between the RE atoms~\cite{Ormaza_NanoLett2016,BlancoRey2022_PRR} and a similar Curie temperature of $T_C$= 19\,K and 22\,K, respectively.  However, these materials differ in their magnetic anisotropic behavior, which is defined by the RE atom.  HoAu$_2$ has an out-of-plane (OOP) easy-axis of magnetization, while GdAu$_2$ has an in-plane (IP) one~\cite{Cavallin2014,Fernandez20Nanoscale,Ormaza_NanoLett2016}. In the case of HoAu$_2$ the main anisotropy contribution arises from localized 4$f$ electrons. On the contrary, in the case of GdAu$_2$, as Gd has an $L$ = 0, the magnetocrystalline anisotropy originates from specific spin-orbit splittings that take place in the RE($d$)-Au($s$) hybrid bands of the GdAu$_2$ band structure~\cite{BlancoRey2022_PRR}. In this work, we explore the magnetic anisotropy of the spinterfaces formed by the adsorption of  CuPc on HoAu$_2$ and GdAu$_2$ MLs, finding that the inherent OOP anisotropy of the CuPc molecule is enhanced upon adsorption on the HoAu$_2$ substrate, while it is attenuated on GdAu$_2$.

In order to characterize comprehensively our hybrid interfaces and the effect of the Gd and Ho atoms on their magnetic properties, we resort to X-ray absorption spectroscopy (XAS), X-ray magnetic circular dichroism (XMCD)
and multiplet simulations. The CuPc/REAu$_2$ systems emerge as excellent candidates for this investigation, due to the weak chemical interaction of CuPc with the mentioned substrates GdAu$_2$ and HoAu$_2$~\cite{Castrillo2023}. Hence, we can discard strong hybridization effects and new spin-polarized
interfacial states as the source for the observed robust and anisotropic AFM exchange coupling.

\section{Results and discussion}

\begin{figure*}[bt!]
\centerline{\includegraphics[width=1.5\columnwidth]{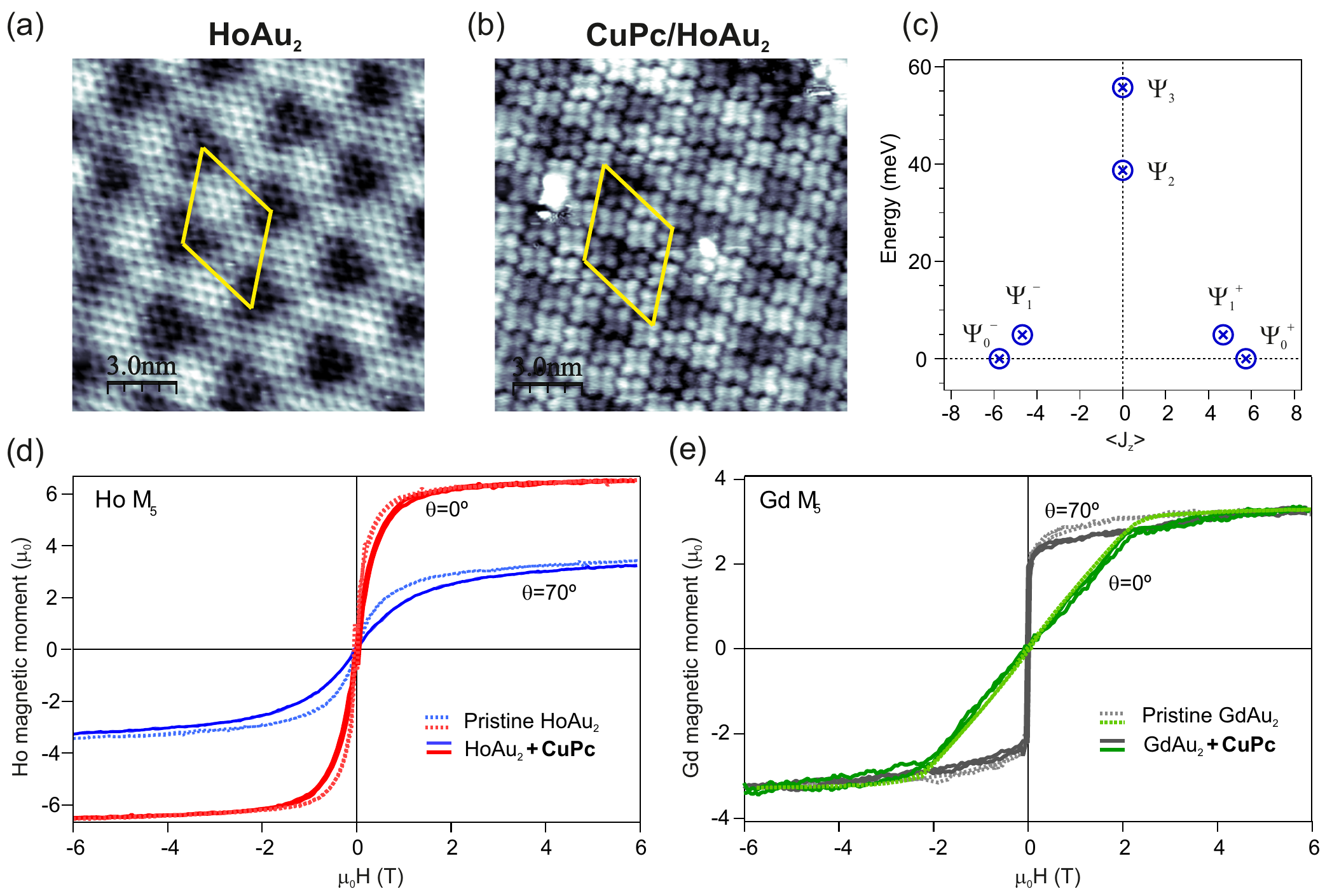}}
\caption{STM images at room temperature of (a) HoAu$_2$ ($I$ = 0.3 nA; $U$ = 0.7 V) and (b) 1 ML CuPc/HoAu$_2$ ($I$ = 0.2 nA; $U$ = 0.9 V) revealing the atomically resolved structure of the investigated systems. The solid rhombo in yellow denotes de underlying Moir{\'e} lattice of the REAu$_2$ surface compound. Image sizes are (50$\times$50) nm$^2$ . (c) The energy splitting of the quantum levels of the Ho$^{3+}$ (4f) orbital in the HoAu$_2$ ML obtained from the multiplet calculations. The ground state $\Psi_0$ and the first excited state $\Psi_1$ are doublets with unquenched $\langle J_z \rangle$. (d), (e) XMCD magnetization curves measured at 4\,K on pristine REAu$_2$ and on CuPc/REAu$_2$ surfaces, RE=Ho, Gd, respectively.
The loops were measured in OOP ($\theta$=0$^{\circ}$) and IP geometry ($\theta$=70$^{\circ}$) at
the Ho and Gd M$_5$ absorbtion edge.}
\label{fig:XMCDloops}
\end{figure*}

Monolayers (MLs) of HoAu$_2$ and GdAu$_2$ are used as substrates for the adsorption of CuPc molecules. Figure~\ref{fig:XMCDloops} (a) and (b) display STM images of the pristine HoAu$_2$ surface and the same covered by 1ML of CuPc. The latter is characterized by a flat and commensurated growth on top of the REAu$_2$ surface compounds~\cite{Castrillo2023}. Moreover, upon CuPc adsorption the representative Moir\'e lattice of the REAu$_2$ surfaces, formed by the lattice mismatch between Au(111) and  and REAu$_2$ layers is still visible. We investigate the magnetic properties of the formed hybrid metal-organic interfaces CuPc/REAu$_2$ (RE = Gd, Ho) by  XAS and XMCD. With that in mind, 0.7ML of CuPc was grown on top of both surface compounds. Spectra were recorded with circularly polarized light at the Cu L$_{2,3}$ and RE M$_{4,5}$  absorption edges in both out-of-plane (OOP) and in-plane (IP) geometries. In OOP, both magnetic field and X-rays are perpendicular ($\theta=0^{\circ}$) to the surface, while in IP the magnetic field and the X-rays are nearly parallel ($\theta=70^{\circ}$) to the surface. Additionally, X-ray linear dichroism (XLD) measurements were carried out on the adsorbed CuPc to investigate its orientation with respect to the plane of the substrate (see Methods).

\subsection{Magnetic robustness of REAu$_2$ monolayers upon CuPc adsorption}
\label{sec:change_in_subs}

GdAu$_2$ and HoAu$_2$ MLs display IP and OOP magnetic anisotropy, respectively, as determined by XMCD measurements~\cite{BlancoRey2022_PRR,Fernandez20Nanoscale}. The IP anisotropy of GdAu$_2$ arises from a band anisotropy,
extensively explored by DFT calculations~\cite{BlancoRey2022_PRR}. In contrast, HoAu$_2$ displays an OOP anisotropy that results from the $^5H_8$ configuration
of the Ho $4f$ orbital, as previously determined~\cite{Fernandez20Nanoscale}.
Before turning our attention to the effect of the molecular layer, we delve further into the
HoAu$_2$ anisotropy by performing multiplet simulations of the experimental XMCD spectra.
To this end, we have used a single-ion Hamiltonian
that includes on the same footing Coulomb, spin-orbit and Zeeman terms, together with a
3-fold symmetric crystal field (CF).
The methodology is described in the Methods section and the
Supporting Information~\cite{bib:SM} sections~\ref{SM-sec:methods_multiplet} and \ref{SM-sec:ho_spectra_fit}.
Figure~\ref{fig:XMCDloops}(c) shows the lowest-lying energy levels of the Ho $J=8$ multiplet
obtained after fitting the CF parameters to the experimental spectra at $\theta = 0^\circ$ and $70^\circ$ incidences.
We find that the ground state $\Psi_0$ is a doublet with $\langle J_z \rangle = \pm 5.8$ and the first excited
state $\Psi_1$ is another doublet with $\langle J_z \rangle = \pm 4.7$, with an energy difference of 4.9\,meV
between them. For the second excited state $\Psi_2$, which lies at a high energy of 38.7\,meV,
the $\langle J_z \rangle$ value becomes quenched by the quantum tunneling allowed by the
3-fold symmetry of the CF. This energy value is a measure of the strong OOP magnetic anisotropy of this system.
This scenario of symmetry-protected magnetic bistability is qualitatively similar
to that of Ho adatoms on Cu(111) \cite{Donati2014_PRL}, with the difference that
the $\langle J_z \rangle$ quenching occurs for an energy larger by one order of magnitude in the
HoAu$_2$ case, which benefits the bistability preservation.
The reason for this difference in the level splittings is the intense CF
felt by the Ho$^{3+}$ ion embedded in the two-dimensional alloy structure (see Table~\ref{SM-tab:hoau2_CFfit}).

Upon CuPc adsorption, the XMCD magnetization curves of both REAu$_2$ MLs suffer small changes that consist mainly in a slight reduction of the magnetization signal, but without significant anisotropy variations. A similar behavior has been reported for organic nanowires grown on
GdAu$_2$ surfaces~\cite{Abadia2017_ACSNano}, where a decrease of $\sim5$\,K in  the Curie temperature ($T_C$) was detected as well. The XMCD magnetization is proportional to the magnetic moment of the sample, but as it is element sensitive, this value gives the magnetic response from the measured element, i.e. from the RE atoms. Fig.~\ref{fig:XMCDloops} displays the XMCD loops measured at the Gd
and Ho M$_5$ absorption edges at IP and OOP geometries, before and after adsorption of 0.7 ML of CuPc.
As we know from previous studies, CuPc molecules grown on HoAu$_2$ and GdAu$_2$ surfaces form densely packed physisorbed
monolayers~\cite{Castrillo2023}. Therefore, we do not expect that the CuPc ML significantly alters
the energy levels of the Ho$^{3+}$ multiplet (Fig.~\ref{fig:XMCDloops} (c)) or the $J=7/2$ multiplet of Gd$^{3+}$.
Under this assumption, the differences observed in the REAu$_2$ XMCD
loops upon CuPc physisorption can be attributed to
(i) the presence of a Cu-RE magnetic exchange interaction and
(ii) subtle changes in the RE-RE RKKY interaction.
The first effect, namely the Cu-RE exchange coupling, will be analyzed and discussed in detail below using the Cu L$_3$
intensities measured on CuPc/REAu$_2$ interfaces. Concerning the second effect,
we recall that the RKKY exchange is mediated by itinerant electrons. To model the RKKY interaction, we have considered an effective magnetic field $B_\textnormal{eff}$ acting on the RE atoms with $B_\textnormal{eff}=B + \hat\gamma \cdot \langle \vec\mu_{J,\mathrm{RE}} \rangle$. The first term $B$ represents the external field and the second term is a self-consistent field (a Weiss-like field) that accounts for the exchange interactions between the expected total magnetic moments $\langle \vec\mu_{J,\mathrm{RE}}\rangle$ of the RE atoms. The $\hat\gamma$ tensor is anisotropic, in the sense that its components differ for magnetic moments aligned parallel or perpendicular to the surface (see section~\ref{SM-sec:magcurve_model_subs} of ~\cite{bib:SM} for a numerical analysis of $\hat\gamma$ in bare REAu$_2$ substrates). In a previous study, we detected a non-negligible impact of the CuPc adsorption on the electronic structure of HoAu$_2$ and GdAu$_2$ MLs~\cite{Castrillo2023}. In fact, CuPc is physisorbed on these surfaces, but there is an energy shift of the RE$(d)$-Au$(s)$ surface states of about 35 meV towards the Fermi level upon CuPc adsorption. The qualitative net effect of CuPc physisorption is a reduction of the self-consistent field in both substrates, i.e., an attenuation of the RKKY coupling, which may imply a reduction of the Curie temperature.This behaviour is consistent with the modification of magnetization curves upon CuPc adsorption observed in Fig.~\ref{fig:XMCDloops}(d).

\subsection{CuPc magnetic anisotropy on REAu$_2$ monolayers}
\label{sec:change_in_mol}

\begin{table*}[tb!]
\caption{\label{tab:sumrules} Magnetic moments $\mu_{L}$ and $\mu_S^\textnormal{eff}$  of
CuPc/HoAu$_2$ and CuPc/GdAu$_2$ and the magnetic moment ratios between the OOP
($\theta$=0$^{\circ}$) and the IP geometry ($\theta$=70$^{\circ}$). The values are  obtained from the sum
rules analysis of the XMCD measurements performed at 6\,T. The error range is of 0.005 $\mu_{B}$ for $\mu_{L}$(0$^{\circ}$) and $\mu_{L}$(70$^{\circ}$). For $\mu_S^\textnormal{eff}$(0$^{\circ}$) the error range amounts to 0.05 $\mu_{B}$ and for $\mu_S^\textnormal{eff}$(70$^{\circ}$) to 0.02 $\mu_{B}$. The last row corresponds to the simulated bare CuPc values (see Supporting Information section~\ref{SM-sec:cupc_anis}).
}
   \begin{tabular}{c|c|c|c|c|c|c}
                     & $\mu_{L}$(0$^{\circ}$) ($\mu_{B}$) & $\mu_S^\textnormal{eff}$(0$^{\circ}$) ($\mu_{B}$) & $\mu_{L}$(70$^{\circ}$) ($\mu_{B}$) & $\mu_S^\textnormal{eff}$(70$^{\circ}$) ($\mu_{B}$) & $\mu_{L}$(0$^{\circ}$)/$\mu_{L}$(70$^{\circ}$) & $\mu_S^\textnormal{eff}$(0$^{\circ}$)/$\mu_S^\textnormal{eff}$(70$^{\circ}$)\\
     \hline
         CuPc/HoAu$_2$ & 0.1 & 1.36 & 0.04 & 0.17 & 2.5 & 8.1 \\
         CuPc/GdAu$_2$ & 0.07 & 1.05 & 0.06 & 0.28 & 1.1 & 3.8  \\
         CuPc (theory) & 0.08 & 2.34 & 0.03 & 0.31 & 2.8 & 7.6  \\

   \end{tabular}
\end{table*}

\begin{figure*}[t!]
\centerline{\includegraphics[width=1.9\columnwidth]{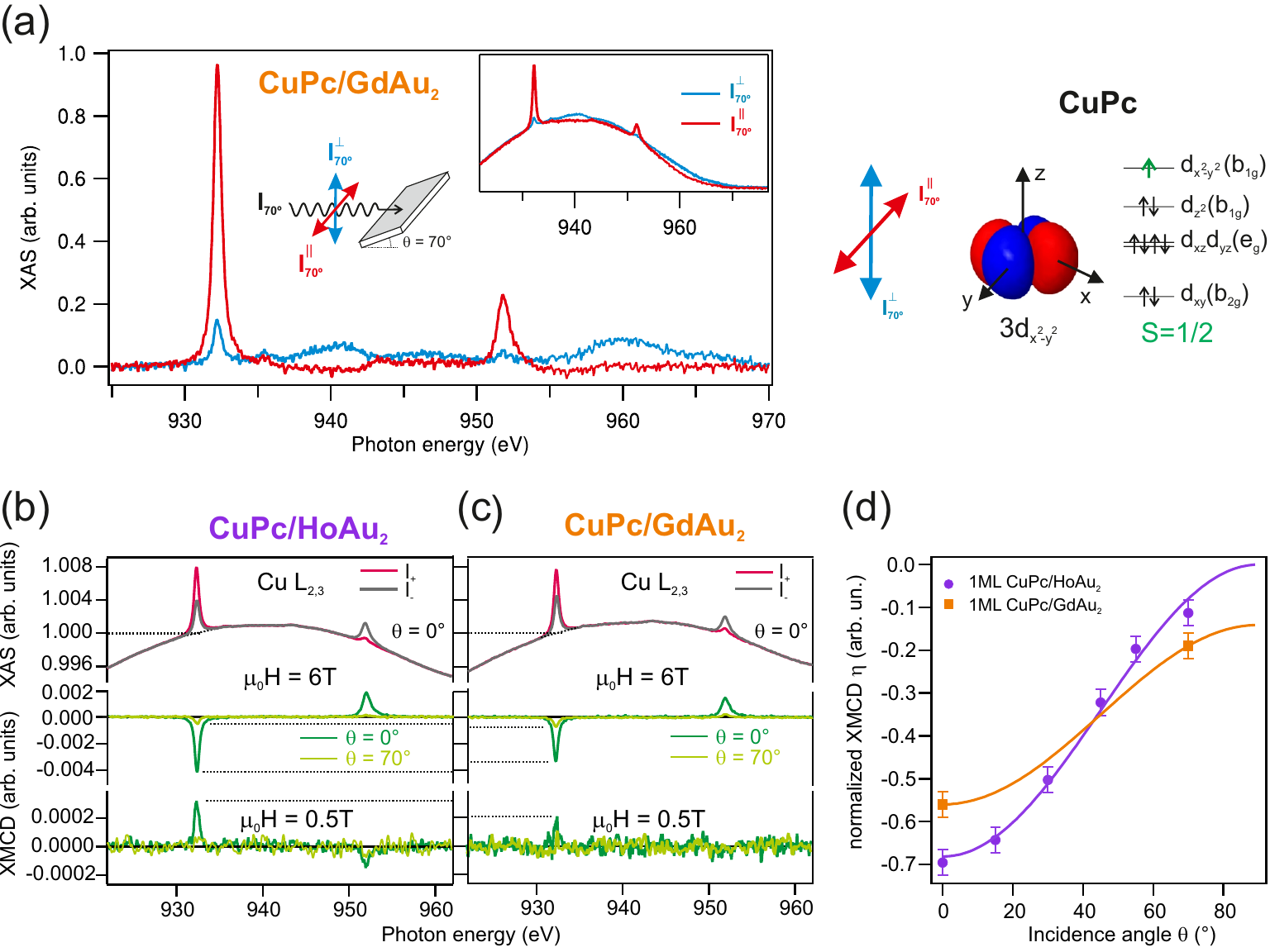}}
\caption{\textbf{}
XLD, XAS, and XMCD spectra measured at the Cu L$_{2,3}$ edge on CuPc/REAu$_2$ samples.
(a): XLD spectra of 1\,ML CuPc on GdAu$_2$ at room temperature. Spectra were recorded with both, vertical (in red) and horizontal (in blue) polarized light at a fixed incident angle of 70$^{\circ}$. The diagram shows the electron filling scheme for CuPc electrons.
(b) and (c): XAS and XMCD of 0.7\,ML CuPc on HoAu$_2$ and GdAu$_2$, respectively measured at 4\,K. The top panels indicate the XAS measurements at an applied field of 6\,T in OOP geometry (normalized to 1 at the base of the L$_3$ peak). The lower panels (in green) reveal the XMCD spectra resulting from the XAS at IP and OOP geometry. XMCD measurements are displayed for applied fields of 6\,T and  0.5\,T.
(d) Angular variation of the Cu L$_3$ normalized XMCD spectra $\eta$ at 6\,T for 1\,ML CuPc/HoAu$_2$ and CuPc/GdAu$_2$.
}
\label{fig:XMCD}
\end{figure*}

X-ray linear dichroism (XLD) measured at the N K (see Supporting Information section S1) and Cu L$_{2,3}$ edges
confirms the planar adsorption of CuPc molecules on REAu$_2$ surface alloys.
Fig.~\ref{fig:XMCD}(a) discloses the Cu L$_{2,3}$ XLD measurements carried out on the CuPc/GdAu$_2$ system. Due to the high background intensity arising from the Au substrate, long range EXAFS fluctuations are induced. Therefore, the copper transition is located on top of a sinus-like curve (see inset) that is subtracted for better visualization. The white line shows a clear linear dichroism with a much larger intensity when $\vec{E}$ is
parallel to the surface plane. The main peak of the XLD pattern is assigned to the
transition from the Cu 2$p_{x,y}$ localized orbital into the singly occupied
molecular orbital (SOMO) Cu 3$d_{x^2-y^2}$ with spin $S=1/2$~\cite{Carniato_PRB}.

Next we measured XAS and XMCD at the Cu L$_{2,3}$ edge at 6\,T and 0.5\,T. Figs.~\ref{fig:XMCD}(b) and (c) show the XAS and XMCD spectra corresponding to CuPc/GdAu$_2$ and CuPc/HoAu$_2$, respectively. An inversion in the sign of the XMCD signal at 0.5\,T is detected, which indicates an antiparallel ordering of the Cu magnetic moment to the underlying REAu$_2$ substrates at low fields. The application of larger applied magnetic fields forces the Cu magnetic moment to rotate in the direction of the field due to the Zeeman effect. In order to study the magnetic anisotropy of the CuPc/REAu$_2$ interfaces, we have performed various XAS/XMCD measurements at different incidence angles of applied magnetic field/light direction. Fig.~\ref{fig:XMCD}(d) shows the angular dependence of the normalized L$_3$ XMCD intensities at 6\,T. The original XAS curves were normalized by a multiplying factor such that the background at the L$_3$ peak has always the same intensity. The normalized XMCD intensity $\eta$ is then calculated as
\begin{equation}
\eta = \frac{I_+ -I_-}{ \frac{I_+ + I_-}{2}-1 } .
\end{equation}
On both substrates, the highest L$_3$ XMCD intensity is measured at OOP geometry ($\theta$=0$^{\circ}$), where the CuPc magnetization is maximum. The same behavior has been observed on CuPc MLs and submonolayers adsorbed on Ag(100) or Au(110) surfaces~\cite{bib:stepanow10,Gargiani_PRB,Bartolome,Mugarza_PRB2012}. This is due to the fact that both the quantum orbital momentum $L_z$ and the spin dipole moment $T_z$ of CuPc show a strong OOP anisotropy~\cite{bib:stepanow10}. In this regard, we have calculated $T_z$ and $L_z$ values for the bare CuPc molecule (see section \ref{SM-sec:cupc_anis} of the Supplementary Material), identifying unequivocally the intrinsic OOP anisotropy character of the free molecule. The strong L$_3$ intensity variation between $\theta$=0$^{\circ}$ and $\theta$=70$^{\circ}$ observed in CuPc/HoAu$_2$ is remarkable and it is shown in Fig.~\ref{fig:XMCD}(d). In the CuPc/GdAu$_2$ system this variation is clearly reduced. A fit of the data points by the angular dependence
\begin{equation}
\eta = \eta_\perp \cos^2\theta + \eta_\parallel \sin^2\theta
\end{equation}
allows to determine the relation $\eta_\perp / \eta_\parallel$ for both systems CuPc/HoAu$_2$ and CuPc/GdAu$_2$~\cite{Gargiani_PRB}.
In the case of the HoAu$_2$ substrate this ratio results to be 11.5, while in the case of GdAu$_2$ is reduced to 4.0. Thus, it is evident that the HoAu$_2$ ML enhances the Cu L$_3$ signal at OOP geometry ($\theta$=0$^{\circ}$), and thus promotes the intrinsic OOP anisotropy of the molecule, while GdAu$_2$ attenuates it. The sum rule analysis allows to extract from the XMCD data measured at 6\,T the $\mu_{L}$ and
$\mu_S^\textnormal{eff}$ values at OOP ($\theta=0^{\circ}$) and IP geometry ($\theta=70^{\circ}$). Table~\ref{tab:sumrules}
summarizes those values and their respective ratios in comparison with the theoretical ones for bare CuPc.
The resulting ratios $\mu_{L}$(0$^{\circ}$)/$\mu_{L}$(70$^{\circ}$) and
$\mu_S^\textnormal{eff}$(0$^{\circ}$)/$\mu_S^\textnormal{eff}$(70$^{\circ}$) confirm the strong OOP anisotropy of
the CuPc/HoAu$_2$ system, and the importance of
the match between anisotropy axes of substrate and molecule
in order to get a more efficient magnetization of the molecular layer.

\subsection{Anisotropic exchange between CuPc and REAu$_2$ substrates}
\label{sec:anisexch}

\begin{figure*}
\includegraphics[width=1.5\columnwidth,angle=0]{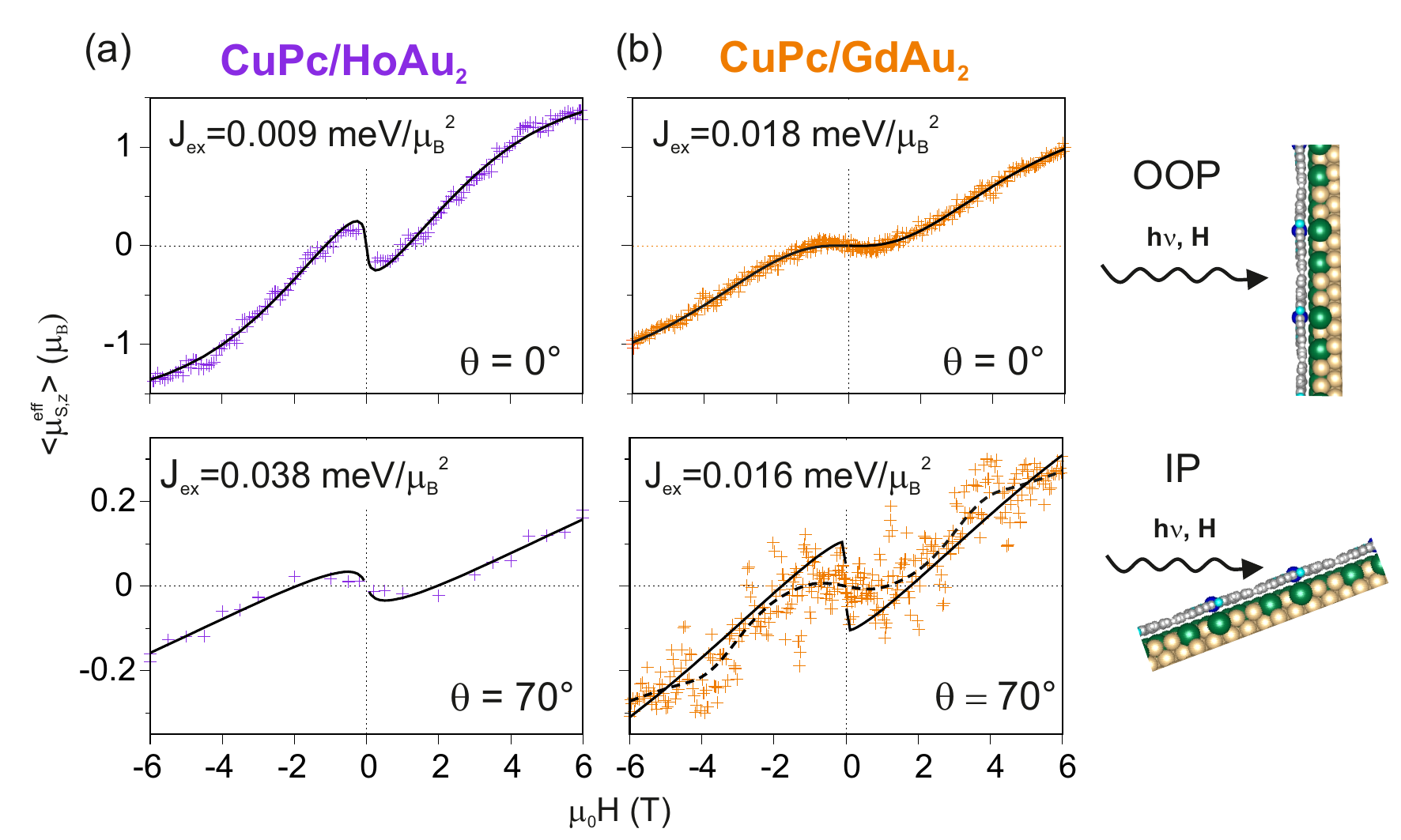}
\caption{\label{fig:XMCDloopsCuPc}
XMCD magnetization curves and simulations at the Cu L$_3$ edge of (a) CuPc/HoAu$_2$, and (b) CuPc/HoAu$_2$ hybrid interfaces. The experimental data measured at 4\,K in OOP ($\theta=0^{\circ}$) and IP ($\theta=70^{\circ}$) geometries are represented in purple (orange) for CuPc/HoAu$_2$ (CuPc/GdAu$_2$). The solid black lines display the theoretical model. In the bottom-right panel,
the dashed line is a smoothed representation of the scattered experimental data to guide the eye.
To compare the data, the experimental intensities are normalized to the experimental
$\langle \mu_S^\textnormal{eff,z} \rangle$ values obtained by sum rules analysis at 6\,T (see Table~\ref{tab:sumrules}).
The labels $\mathcal{J}_{ex}$ indicate the Cu-Ho and Cu-Gd exchange coupling constants
obtained from fits to Eq.~\ref{eq:cupc_curve}.}
\end{figure*}

The XMCD magnetization curves measured at the Cu L$_3$ edge of the CuPc/REAu$_2$ samples are shown in Fig.~\ref{fig:XMCDloopsCuPc}.
These measurements probe the Cu magnetic moment $M_\mathrm{Cu}$.
Both systems display a similar trend, but with particular features that depend on the substrate and the direction of
the applied magnetic field. At OOP geometry and low fields (below 1\,T) the already observed AFM
coupling between Cu and RE atoms is confirmed.
It is worth to note that, in the case of the CuPc/GdAu$_2$ interface, $M_\mathrm{Cu}$ is slightly reduced compared to CuPc/HoAu$_2$. The former is related to the magnetization of the GdAu$_2$ surface, which in OOP geometry has a reduced magnetization at low fields,
as it is seen in the XMCD loops of Fig.~\ref{fig:XMCDloops}(e).
The XMCD magnetization curves measured in IP geometry reveal that between 0\,T and 2\,T
$M_\mathrm{Cu}$ is zero or smaller than the experimental error in both systems. This behavior points out the relatively strong perpendicular
anisotropy of CuPc, which hinders the polarization of $M_\mathrm{Cu}$ in the IP direction on both substrates.
Nevertheless, at higher applied fields there is a change in the magnetization behavior due to the Zeeman interaction that induces the gradual rotation of $M_\mathrm{Cu}$ and allows a larger projection of $M_\mathrm{Cu}$ along the applied field direction and
REAu$_2$ magnetization, although not reaching a full saturation in either IP or OOP directions.

We have modelled the CuPc magnetization curves on both REAu$_2$ substrates neglecting the molecule-molecule
magnetic interactions.
The huge anisotropy of the spin dipole moment $T_z$ and effective spin moment $\mu_S^\textnormal{eff}$ of CuPc
is reflected in a strong attenuation of the XMCD $L_3$ peak intensity at IP geometry.
By treating Cu$^{2+}$ as a $S=1/2$ Zeeman-split two level system, we obtain the expectation value
\begin{equation}
\langle \mu_{S,z}^\textnormal{eff} \rangle (\mu_0 H) = \frac{1}{2} \tanh
   \frac{\mu_{S,z}^\textnormal{eff}(\mu_0 H + \mathcal{J}_{ex} \langle \mu_{J,\mathrm{RE},z}^\textnormal{eff} \rangle )}{k_BT}
\label{eq:cupc_curve}
\end{equation}
Here, the quantities to be adjusted are $\mathcal{J}_{ex}$,
which represents the magnetic exchange between the Cu and RE atoms,
and $\mu_{S,z}^\textnormal{eff}$, which is the Cu saturated effective spin moment projection on the applied field ($\mu_0 H$) direction.
The substrate magnetization as a function of the applied field $\mu_0 H$ enters in the model as the calculated RE expected effective spin moment
projections $\langle \mu_{J,\mathrm{RE},z}^\textnormal{eff} \rangle$, shown in Fig.~\ref{SM-fig:magcurve_subs}.
This way, the model accounts for the anisotropic features of HoAu$_2$ and GdAu$_2$ surfaces.
The experimental data fits to this model and the resulting $\mathcal{J}_{ex}$ values are
included in Fig.~\ref{fig:XMCDloopsCuPc} as solid lines for OOP and IP geometry.
The intensities have been normalized to the $\mu_S^\textnormal{eff}$
values extracted from the sum rules values of Table~\ref{tab:sumrules}.
The field at which $\langle \mu_{S,z}^\textnormal{eff} \rangle$ cancels out is proportional
to the AFM exchange coupling constants $\mathcal{J}_{ex}$ between Cu and RE atoms. The low orders of magnitude of the fitted $\mathcal{J}_{ex}$ ($\sim 0.01$\,meV/$\mu_B^2$)
are consistent with the weak physisorption of CuPc on REAu$_2$~\cite{Castrillo2023} and
similar to those reported in the literature for TbPc$_2$ on Ni substrates~\cite{bib:lodi11}.
The fits show a pronounced anisotropic behavior of the Cu-Ho exchange. The ratio between the exchange constants
for IP and OOP applied fields in this case is $\mathcal{J}_{ex}^\parallel / \mathcal{J}_{ex}^\perp = 4.2$,
whereas the Cu-Gd exchange is more isotropic with $\mathcal{J}_{ex}^\parallel / \mathcal{J}_{ex}^\perp = 1.1$.
Similar strongly anisotropic interactions between metal-organic molecules and ferromagnetic substrates
have also been reported for Cu-tetraazaporphyrin on Fe$_3$O$_2$~\cite{bib:klanke13}
and TbPc$_2$ on oxidized and reduced Ni thin films \cite{bib:lodi11}.
In these systems the coupling follows a superexchange mechanism, where the organic ligands
play a determinant role as intermediate orbitals for
hopping~\cite{bib:goodenough55,bib:goodenough58,bib:kanamori59}.
These superexchange interactions show anisotropy as a consequence of spin-orbit coupling~\cite{bib:moriya60a,bib:moriya60}
\footnote{Moriya's derivation~\cite{bib:moriya60} shows the existence of an
anisotropic exchange contribution of second order in the SOC, which also appears in centro-symmetric systems.
Note, this derivation does not consider orbital degeneracy. See also Ref.~\cite{bib:stevens53}.},
and this anisotropy is facilitated by the presence of states with orbital degeneracies,
a mechanism known as orbital dependent exchange (ODE)~\cite{bib:oguchi65,bib:baker71,bib:dai08,bib:palii11}.
This mechanism is relevant in the case of $3d-3d$ and $4f-3d$ exchange interactions in crystals and molecular
magnets~\cite{Liu,Piquer2015}. In particular, in systems where the RE orbital moment $L$ is
unquenched, ODE is enhanced~\cite{bib:borras01,bib:palii05,bib:palii11,bib:dreiser12}.
Indeed, the analysis of Fig.~\ref{fig:XMCDloopsCuPc} shows that the ODE effect is smaller for
CuPc on GdAu$_2$ ($L=0$) than on HoAu$_2$ ($L=6$).

\section{Conclusion}

We have confirmed the AFM coupling stablished at the hybrid metalorganic interface between a CuPc monolayer and a ferromagnetic single atomic layer of REAu$_2$ (RE=Gd, Ho). The latter is set by a 3$d$-4$f$ superexchange interaction between Cu and RE atoms. Moreover, this AFM coupling across the interface seems to be robust up to approximately 1.5\,T. The natural OOP anisotropy of a CuPc ML is enhanced upon adsorption on the HoAu$_2$ substrate, which shows an OOP easy-axis of magnetic anisotropy.
It is worth noting that CuPc molecules are physisorbed on REAu$_2$ surfaces and do not display strong hybridization or chemical interaction with the substrate. In this context, the aforementioned magnetic properties cannot be ascribed to the generation of a new spin-polarized interfacial state, as it is often the case in strongly hybridized spinterfaces.
The use of RE atoms introduces strong spin-orbit effects that may promote a large anisotropic exchange coupling between the Cu and RE atoms. This is the case of the hybrid spin interface CuPc/HoAu$_2$, which displays an exchange interaction that is four times stronger for magnetic fields applied in IP geometry than in the OOP one, whereas CuPc/GdAu$_2$ exhibits isotropic exchange coupling. We explain this behavior by the orbital dependent exchange mechanism, which appears in systems with an unquenched orbital moment $L$, this is, states with an orbital degeneracy. The effect of anisotropic exchange coupling found in the hybrid interface CuPc/HoAu$_2$ arises as a relevant result for the development of magneto-resistive devices with extended functionalities, allowing a modulation of the coupling strength upon a $90^\circ$ rotation of the external field. In conclusion, the stable AFM coupling at the metal organic interfaces of CuPc/REAu$_2$ can show a tunable anisotropic character, depending on the RE atom of the substrate, which represents a step forward in the field of molecular spintronics.

\section{Methods}


Samples have been prepared in an ultra-high vacuum (UHV) chamber at a base pressure of
2$\times$10$^{-10}$ mbar. Several Au(111) single crystals were used as substrates, that were
cleaned by cycles of Ar$^+$ ion sputtering ($E_{\textnormal{kin}}$ = 1\,keV) and annealing to 500$^\circ$C.
The different RE-Au$_2$ surface compounds were grown \textit{in-situ} by evaporation of
small amounts of RE atoms on the Au(111) surface, held at a fixed temperature. The
optimal growth temperature for the substrate is of 420$^\circ$C for HoAu$_2$ and varies between 400-450$^\circ$C for GdAu$_2$. Below these
temperatures the characteristic Moir\'e was not well formed, and above them the RE metals start to diffuse into the bulk or re-evaporate from the surface. CuPc molecules have been evaporated on
the REAu$_2$ surface with deposition rates of 0.05\,ML/min. The calibration of the CuPc layer thickness was carried out by gradual
depositions of CuPc/Au(111) and the subsequent low energy electron diffraction (LEED) analysis. The coverage of the first visible diffraction spots of the molecules at RT was defined as 0.9 ML~\cite{Stadtmueller2011_PRB}.
The substrate temperature during CuPc evaporation was set to room temperature.

XMCD and XLD experiments were realized at the BOREAS beamline of ALBA synchrotron radiation facility in Spain.
Absorption spectra were acquired in total electron yield at the Cu L$_{2,3}$ as well as Ho and Gd M$_{4,5}$ edges.
XLD spectra were collected at room temperature at a incidence angle of  $70^\circ$, rotating the light polarization
from vertical ($\vec{E}$ in plane) to horizontal ( $\vec{E}$ out of plane). XMCD spectra were acquired at different
photon and magnetic field incidence angles (from $\theta = 0^\circ$ to $\theta = 70^\circ$) in order to study the
magnetic anisotropy of the different interfaces. In out-of-plane (OOP) geometry the light propagation vector and the applied
magnetic field are normal to the sample surface, while in in-plane (IP) geometry the magnetic
field is nearly parallel to the surface.
For the case of CuPc, 99\% of circularly polarized light was used at the Cu L$_{2,3}$ edge, while rare-earth edges were
acquired with 90\% polarization.
XMCD measurements were carried out at or below 4\,K with a variable magnetic field up to $\pm$6\,T.  The XMCD spectrum is the difference between the two x-ray absorption spectroscopy (XAS) spectra recorded with opposite orientation of the magnetic field and/or the circular helicity of the light, which we call $I_+$ and $I_-$ for simplicity. The XMCD signal is proportional to the projection of the magnetization in the direction of the applied magnetic field. Element sensitive magnetization loops were measured by sweeping the photon energies corresponding to the maximum of the XMCD asymmetry signal at the Cu L$_3$, Ho and Gd M$_5$ absorption edges and a pre-edge energy as a function of the magnetic field. The latter was used for normalization and accounts for possible instabilities of the light entering the experimental station. Sum rules were used to obtain orbital $\mu_L$ and effective spin $\mu_S^\textnormal{eff}$ moments of CuPc at $\mu_0$H = 6\,T~\cite{Thole1992_PRL,Carra1993}.
The effective spin moment $\mu_S^\textnormal{eff}$ refers to the measurable value of the expectation values of total atomic spin ($S$) and spin dipole  $T_z$ operators with $\mu^S_\textnormal{eff}$= 2$S$ + 7$T_z$. Both orbital and effective dipole momenta depend on the number of electron holes $n$ in the $d$ and $f$ shell. For the elements considered here Cu, Ho, and Gd, we set as electron holes $n$ = 1 in the Cu $d$-shell and $n$ = 4 and 7 for Ho and Gd holes in the $f$-shell, respectively.
The sum $\mu_L$ + $\mu_S^\textnormal{eff}$ was used to normalize the XMCD magnetization curves.


For a given $4f^n$ occupancy of the Ho ion
the XAS and XMCD M$_{4,5}$ absorption edges spectra of the bare HoAu$_2$ substrate are
simulated with the code Xclaim \cite{bib:xclaim}
by considering electronic transitions $3d \rightarrow 4f$
in a model Hamiltonian for the $4f$ orbital,
where hybridization of $4f$ electrons is neglected.
In this work, the crystal field parameters are optimized
to match the simulated and experimental lineshapes
of the $I_+,I_-$ measured intensities at normal and grazing ($70^\circ$ off-normal)
incidences. Four data sets are simulateously fit in
an automated procedured, which is described in the Supplementary
Material section~\ref{SM-sec:methods_multiplet}
along with further details of the hamiltonian parameterization.
From the hamiltonian eigenergies and eigenstates the
expectation values $\langle J_z \rangle$ of the Ho$^{3+}$
total magnetic moment projection on the field direction are calculated,
which provide a model of the HoAu$_2$ magnetization curves.
Similarly, using suitable crystal field parameters \cite{Rosa_1994}, a single ion hamiltonian for Cu$^{2+}$ in CuPc
has been written and diagonalized (see Supplementary Material section~\ref{SM-sec:cupc_anis}).

\begin{acknowledgments}

This work was financial supported by Spanish Ministerio de Ciencia e Innovaci\'on
(grants No. PID2019-103910GB-I00, PID2019-107338RB-C65,
PID2020-116093RB-C44, and PID2021-123776NB-C21, PID2022-137685NB-I00, PID2022-140845OB-C63, CEX2021-001214-S, and CEX2020-001039-S
funded by MCIN/AEI/10.13039/501100011033/ and by ``ESF investing in your future``),
as well as the Department of Education, Universities and Research of the
Basque Government (grants No. IT1591-22 and IT1527-22 and IKUR Quantum program) and the CERCA Programme/Generalitat de Catalunya and
Diputaci\'on Foral de Gipuzkoa through QUAMOS project within the Gipuzkoa Quantum program.
L.F. acknowledges funding from the European Union's Horizon 2020 research and innovation
programme through the Marie Sk\l odowska-Curie Grant Agreement MagicFACE No. 797109.
Computational resources were provided by DIPC.
Javier Fern\'andez-Rodr\'{\i}guez is acknowledged for useful discussions.

\end{acknowledgments}


\bibliography{magnetCuPc}

\makeatletter\@input{aux4_SM_magnetCuPc_REAu2_200224.tex}\makeatother

\end{document}



\title{Supporting Information  for: ``The role of rare-earth atoms in the anisotropy and antiferromagnetic exchange coupling at a hybrid metal-organic interface''}

\newcommand{\ALBA}[0]{{
ALBA Synchrotron Light Source, carrer de la Llum 2-26, 08290 Cerdanyola del Vall{\`e}s, Spain}}
\newcommand{\CFM}[0]{{
Centro de F\'{\i}sica de Materiales CSIC-UPV/EHU-Materials Physics Center, 20018 San Sebasti\'an, Spain}}
\newcommand{\UPVPMA}[0]{{
Departamento de Pol\'imeros y Materiales Avanzados: F\'isica, Qu\'imica y Tecnolog\'ia,
Universidad del Pa\'is Vasco UPV/EHU, 20018 San Sebasti\'an, Spain}}
\newcommand{\DIPC}[0]{{
Donostia International Physics Center, 20018 Donostia-San Sebasti\'an, Spain}}
\newcommand{\UPVFA}[0]{{
Departamento de F\'isica Aplicada I,
Universidad del Pa\'is Vasco UPV/EHU,  20018 San Sebasti\'an, Spain}}
\newcommand{\IOMCNR}[0]{{
IOM-CNR, Laboratorio TASC, Strada Statale 14 Km 163.5, I-34149 Trieste, Italy}}
\newcommand{\Elettra}[0]{{
Elettra - Sincrotrone Trieste S.C.p.A., Strada Statale 14 Km 163.5, I-34149 Trieste, Italy}}
\newcommand{\TUDD}[0]{{
Institut f\"ur Festk\"orper- und Materialphysik, Technische Universit\"at
Dresden, D-01062 Dresden, Germany}}
\newcommand{\ICN}[0]{{
Catalan Institute of Nanoscience and Nanotechnology (ICN2), CSIC and BIST, Campus UAB, 08193 Barcelona, Spain}}
\newcommand{\ICREA}[0]{{
ICREA-Instituci\'o Catalana de Recerca i Estudis Avan\c{c}ats, Lluis Companys 23, 08010 Barcelona, Spain}}
\newcommand{\IMDEA}[0]{{
Instituto Madrile\~no de Estudios Avanzados, IMDEA Nanociencia, Calle Faraday 9, 28049 Madrid, Spain}}
\newcommand{\Chalmers}[0]{{
Chalmers University of Technology, G{\"o}teborg, Chalmersplatsen 4, 412 96 G{\"o}teborg, Sweden}}
\newcommand{\Nanogune}[0]{{
CIC nanoGUNE-BRTA, 20018 San Sebasti\'an, Spain}}

\author{Mar\'{\i}a Blanco-Rey}
\affiliation{\UPVPMA}
\affiliation{\DIPC}
\affiliation{\CFM}
\author{Rodrigo Castrillo}
\affiliation{\CFM}
\author{Khadiza Ali}
\affiliation{\DIPC}
\affiliation{\Chalmers}
\author{Pierluigi Gargiani}
\affiliation{\ALBA}
\author{Max Ilyn}
\affiliation{\CFM}
\author{Michele Gastaldo}
\affiliation{\ICN}
\affiliation{J. Heyrovsk{\'y} Institute of Physical Chemistry, Czech Academy of Sciences, Dolej{\v s}kova 2155/3, Prague, 18223, Czech Republic }
\author{Markos Paradinas}
\affiliation{\ICN}
\affiliation{Institut de Ci{\`e}ncia de Materials de Barcelona, ICMAB-CSIC, Campus UAB, 08193 Bellaterra, Spain}
\author{Miguel A. Valbuena}
\affiliation{\ICN}
\affiliation{\IMDEA}
\author{Aitor Mugarza}
\affiliation{\ICN}
\affiliation{\ICREA}
\author{J. Enrique Ortega}
\affiliation{\UPVFA}
\affiliation{\CFM}
\affiliation{\DIPC}
\author{Frederik Schiller}
\affiliation{\CFM}
\affiliation{\DIPC}
\author{Laura Fern\'andez}
\affiliation{\Nanogune}
\affiliation{\CFM}
\email{lauraisabel.fernandez@ehu.es}

\date{\today}
\maketitle


\section{Planar adsorption of CuPc molecules on REAu$_2$ surfaces}
\label{sec:exper1}

\begin{figure}[hbt!]
\centerline{\includegraphics[width=0.5\columnwidth]{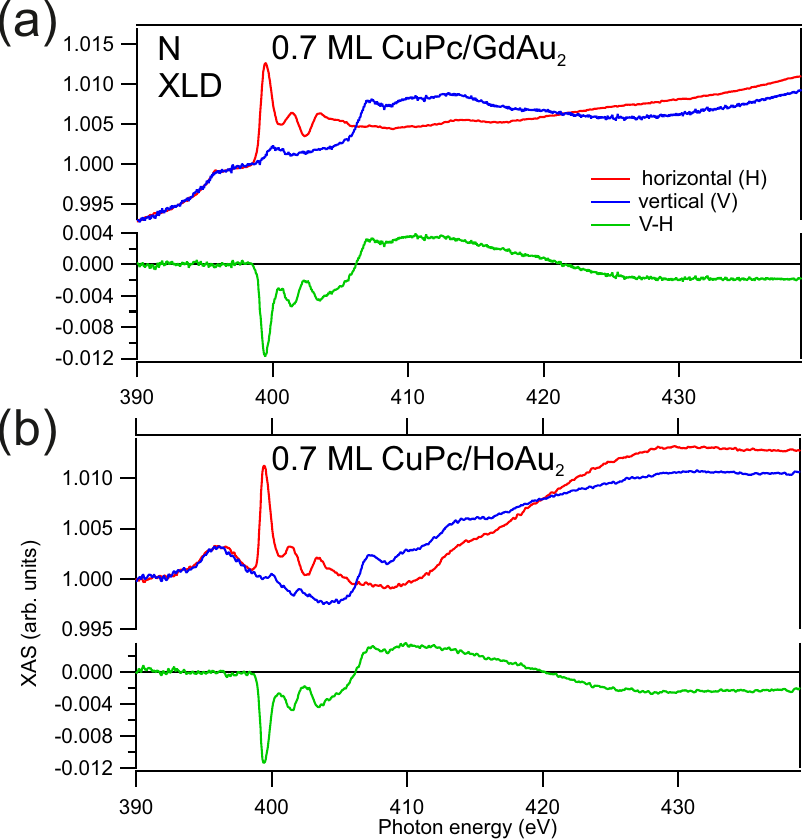}}
\caption{\label{XLD}
XLD at the N L$_{2,3}$ edge at RT on (a) CuPc/GdAu$_2$ and (b) CuPc/HoAu$_2$ samples. Spectra were recorded with both, vertical (in red) and horizontal (in blue) polarized light at a fixed incident angle of 70$^{\circ}$. The green one is the difference between the signals measured with horizontal and vertical polarized light.}
\end{figure}

X-ray linear dichroism (XLD) measurements at the
N K edge (see Fig.~\ref{XLD}) certainly confirm the
planar adsorption growth of CuPc molecules on the REAu$_2$
surface alloys.

\section{Angular dependence of the L$_3$ XAS and XMCD intensities}
\label{sec:exper2}

\begin{figure}[hbt!]
\includegraphics[width=0.9\columnwidth,angle=0]{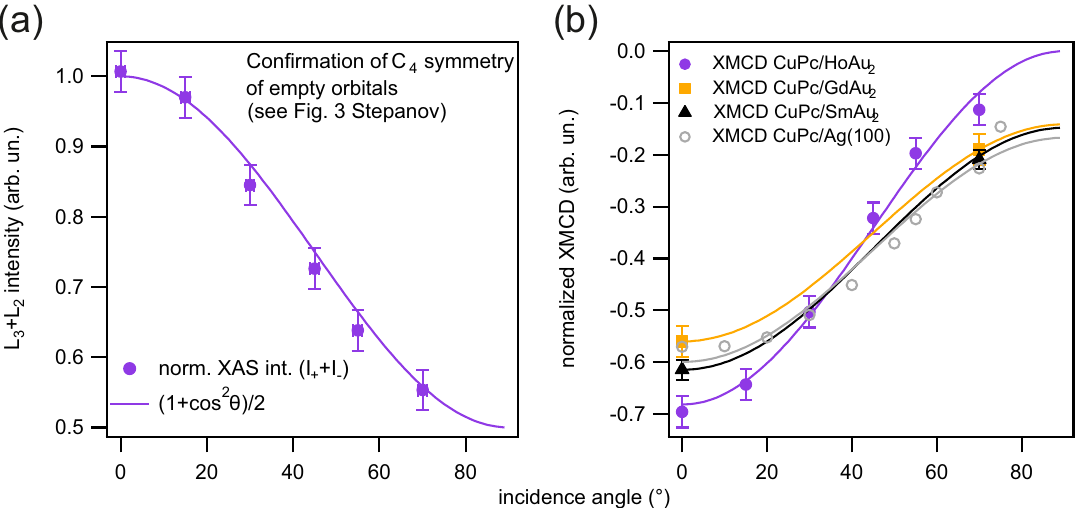}
\caption{\label{C4_L3}
(a) Sum of L$_3$ and L$_2$ peak XAS values as a function of $\theta$. For orbitals having a C$_4$ symmetry rotation axis perpendicular to the substrate,
$I_{\theta}^{+}$ +$I_{\theta}^{-}$
is expected to follow a function $f(\theta)=(1+cos^2 \theta)/2$. Note that, in order to compare spectra recorded at different angles and compensate for changes in the total electron yield, the raw data were first normalized to a common intensity value~\cite{Stepanow_PRB2010_CuPc_Ag}; (b) normalized L$_3$ XMCD peak values as a function of $\theta$ for CuPc adsorbed on HoAu$_2$, GdAu$_2$, Au(100) and multilayers of CuPc on SmAu$_2$.}
\end{figure}

As for CuPc layers on Ag(100) or Au(110)~\cite{Gargiani_PRB, Stepanow_PRB2010_CuPc_Ag}, also in the case of CuPc on REAu$_2$
surface compounds, the molecular orbital associated to the  Cu L$_{2,3}$ transition presents a $C_4$ symmetry, see Fig.~\ref{C4_L3}(a). Note that for orbitals having a C$_4$ symmetry rotation axis perpendicular to the substrate,
$I_{\theta}^{+}$ +$I_{\theta}^{-}$
is expected to follow a function $f(\theta)=(1+cos^2 \theta)/2$,
as given by the mixing of two orthogonal linearly polarized components, one in plane and the other tilted by $\theta$ with respect to the molecular plane.

Fig.~\ref{C4_L3}(b) displays the angular dependence of the $L_3$ XMCD intensity peak measured at the Cu$_{L_2,3}$ edge for CuPc/HoAu$_2$ and CuPc/GdAu$_2$ at 6\,T. The normalized $L_3$ XMCD intensities are plotted as a function of the applied magnetic field/light incidence angle. For comparison we add the normalized values of a 5 ML thick CuPc film, which represents a thick CuPc film that should not have a strong influence of the substrate. Note that these
XAS and XMCD values are very similar to those of the submonolayer of CuPc grown on nonmagnetic Ag(100)~\cite{Stepanow_PRB2010_CuPc_Ag}. For
normalization purposes the original XAS curves
were multiplied by a factor in a way that the background at the $L_3$ peak has always the same intensity. The normalized XMCD intensity $\eta$ is the integral value of the intensities $I_+$ ($I_-$) of the L$_3$ peak obtained for right (left) circularly polarized light applying $\eta = 2\cdot(I_+ -I_-)/(I_+ +I_-)$.

\pagebreak


\section{Multiplet simulation methods}
\label{sec:methods_multiplet}


For a given $4f^n$ occupancy of the Ho ion
the XAS and XMCD $M_{4,5}$ absorption edges spectra are
simulated by considering electronic transitions $3d \rightarrow 4f$
in the model hamiltonian for the $4f$ orbital
\begin{equation}
\mathcal{H} = \mathcal{H}_{Coulomb} - \lambda \mathbf{L}\cdot\mathbf{S} + \mathcal{H}_{CF} +
\mu_B (\mathbf{L} + 2 \mathbf{S}) \cdot \mathbf{B}
\label{eq:Hsia}
\end{equation}
where the terms account for the intraatomic Coulomb interactions,
spin-orbit coupling (SOC), crystal field (CF) potential and Zeeman term, respectively.
The latter term can be also expressed as
$H_Z = \mu_B g_J \mathbf{J} \cdot \mathbf{B}$,
where the external field $\mathbf{B}$ acts on the total magnetic moment and $g_J$ is the Land\'e factor
($g_J=4/5$ for Ho$^{3+}$).
Hybridization of $4f$ electrons is neglected, as ARPES experiments show that
$4f$ bands in HoAu$_2$ are non-interacting \cite{Fernandez20Nanoscale}.
Each setting of parameters defines a hamiltonian for the $4f$ orbitals and thus determines
the many-electron ground state (GS) and excited states (ES).
These are obtained using the computer code Xclaim \cite{bib:xclaim},
together with the simulated XAS and XMCD spectra
by applying the Fermi golden rule to the hamiltonian eigenstates after
broadening the energy levels.
The CF parameters are optimized by trial and error
to match the simulated and experimental lineshapes.

In this work we express the CF parameters in the Wybourne ($B_{kq}$)
and Stevens ($B^k_q$) conventions \cite{bib:newmanngbook}.
Other parameters used in the multiplet simulations (standard
atomic Hartree-Fock data \cite{bib:cowanbook} and
characteristic energies in the lineshapes) are given Table~\ref{tab:hoau2_slater}.
Note that, in order to diagonalize the hamiltonian numerically, a small
magnetic field is needed. We use 0.01\,T.

\begin{table}[h]
\caption{\label{tab:hoau2_slater}
Atomic parameters used in the multiplet simulations for HoAu$_2$:
for the initial and final states in the transition, Slater integrals $F^k$ and $G^k$, and
atomic spin-orbit coupling (SOC) parameters $\xi$, taken from Ref.~\cite{bib:cowanbook}.
Additionally, reduction factor with respect to the Hartree-Fock value.
position $E_0$ of the $M_5$ edge, and smearing widths $\sigma$ applied
to the $M_{4,5}$ peaks of the simulated spectra.
Items marked with $(^\star)$ have been optimized in the lineshape fitting.
All energies are given in eV.}
\begin{ruledtabular}
\begin{tabular}{ccc}
   Ground state  & Final state  &  \\
$F^2(4f,4f) = 15.706 $   &  $F^2(4f,3d) = 10.716 $  &  reduction 85.5\%$^\star$  \\
$F^4(4f,4f) = 9.854  $   &  $F^4(4f,3d) = 5.051 $   &  $E_0 = 1335.42^\star $  \\
$F^6(4f,4f) = 7.089  $   &  $G^1(3d,4f) = 7.799 $   &  $\sigma_{M_5} = 0.3^\star $  \\
$\xi(4f) = 0.273 $       &  $G^3(3d,4f) = 4.576 $   &  $\sigma_{M_4} = 0.5^\star $ \\
                         &  $G^5(3d,4f) = 3.162 $   &  \\
                         &  $\xi(4f) = 0.307  $     &  \\
                         &  $\xi(3d) = 15.07^\star $  &  \\
\end{tabular}
\end{ruledtabular}
\end{table}

In the HoAu$_2$ case the spectral lineshapes are
governed by the CF to a great extent, as proven by the
the strong dependence on the incidence direction.
Taking advange of this observation, we use $c+$ and $c-$ measured intensities instead of
XAS and XMCD curves in the analysis. Thus, for a given set of the other parameters,
we make an automated search of the CF parameters to minimize a cost function
that \emph{simultaneously} fits the four sets of data, namely the $c+$ and $c-$ spectra
obtained at normal and grazing (at $70^\circ$ off-normal) incidences.
This minimization is carried out by simulated annealing and simplex algorithms \cite{bib:nr},
where a cost function $\chi^2_{LD}$ is defined in terms of the logarithmic derivatives
of the energy vs. intensity curves $I(E)$:
\begin{equation}
L(E) = \frac{1}{I(E)}\frac{\mathrm{d}I}{\mathrm{d}E}
\end{equation}
as follows:
\begin{equation}
\label{eq:chi2LD}
\begin{split}
\chi^2_{LD} = \sum_i & [L^{\perp}_e(c^+,E_i)-L^{\perp}_t(c^+,E_i) ]^2  +
                  [L^{\perp}_e(c^-,E_i)-L^{\perp}_t(c^-,E_i) ]^2  +  \\
                     & [L^{\parallel}_e(c^+,E_i)-L^{\parallel}_t(c^+,E_i) ]^2 +
                  [L^{\parallel}_e(c^-,E_i)-L^{\parallel}_t(c^-,E_i) ]^2
\end{split}
\end{equation}
Here $L^{\perp (\parallel)}$ are the logarithmic derivatives
of the spectra for normal (grazing) incidences calculated numerically for each curve,
the subindex $e (t)$ indicates experimental (theoretical) spectra,
$c^+ (c^-)$ indicates right (left) handed light spectra,
and $E_i$ run over the discrete energy data points at the regions
of the spectrum where the XAS peaks lie.
The rationale behind the use logarithmic derivatives in the cost function is to
account numerically for the coincidence of maxima, minima, and elbows in the spectra,
rather than compare bare absolute intensities directly,
in the same spirit as R-factors are used in the analysis of low-energy
electron diffraction intensities \cite{bib:pendry80}.
In order to compare individual lineshape intensities, the spectra are first normalised
by setting the maximum intensity out of the four theoretical
and experimental spectra to a value equal to one.
The magnetic field acting on the total quantum angular momentum $\mathbf{J}$
is an effective field contributed by the applied external field $\mathbf{B}$
and the internal exchange field due to interactions between the Ho magnetic moments
in the HoAu$_2$ monolayer. The XAS lineshapes will be fitted without the
internal field. This contribution will be determined later on by fitting the
magnetization curves of Ho$^{3+}-4f$.

Once the CF parameters are optimized, the expected total magnetic moment of Ho$^{3+}$
projected on the applied field direction at the experimental temperature $T=4$\,K is given by
\begin{equation}
\langle \mu_z^J \rangle = \frac{1}{Z} \sum_i \mu_z^J e^{-E_i/k_BT}
\label{eq:Jzave_magcurve}
\end{equation}
where $Z$ is the partition function and the sum runs over the eigenstates and eigenergies
obtained from diagonalization of the hamiltonian Eq.\ref{eq:Hsia}.


\section{X-ray spectral analysis of the $\mathrm{HoAu}_2$ substrate: $\mathrm{Ho}^{3+}$ multiplets}
\label{sec:ho_spectra_fit}

\begin{figure}
\includegraphics[width=0.45\columnwidth,angle=0]{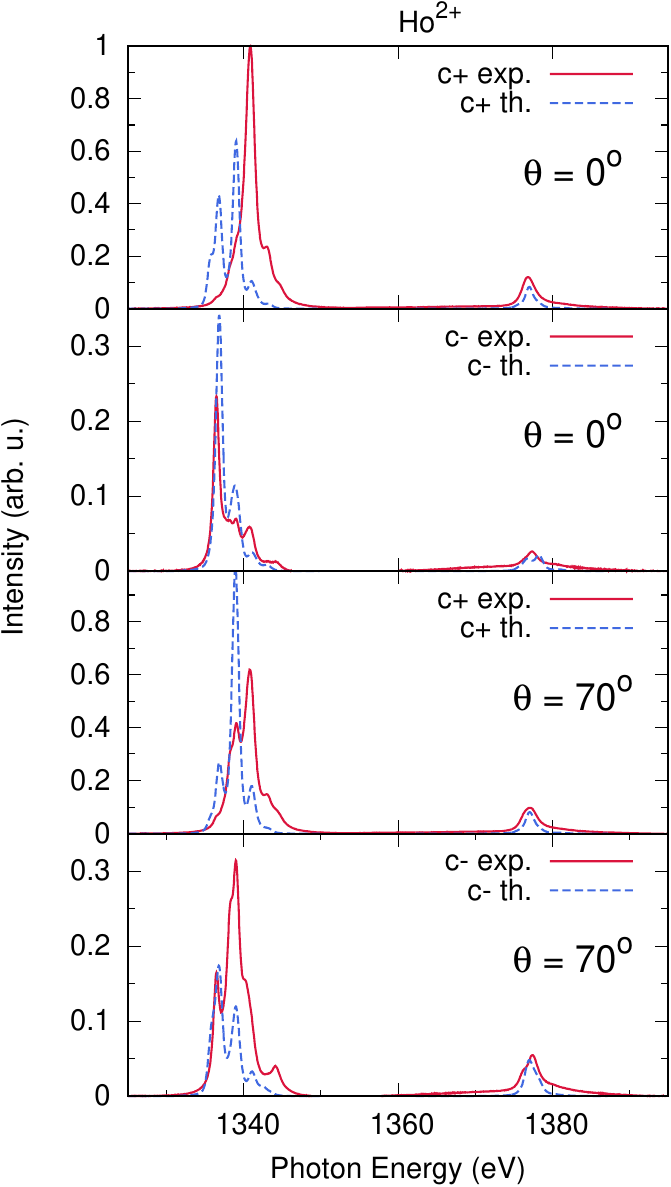}
\hspace{0.2cm}
\includegraphics[width=0.45\columnwidth,angle=0]{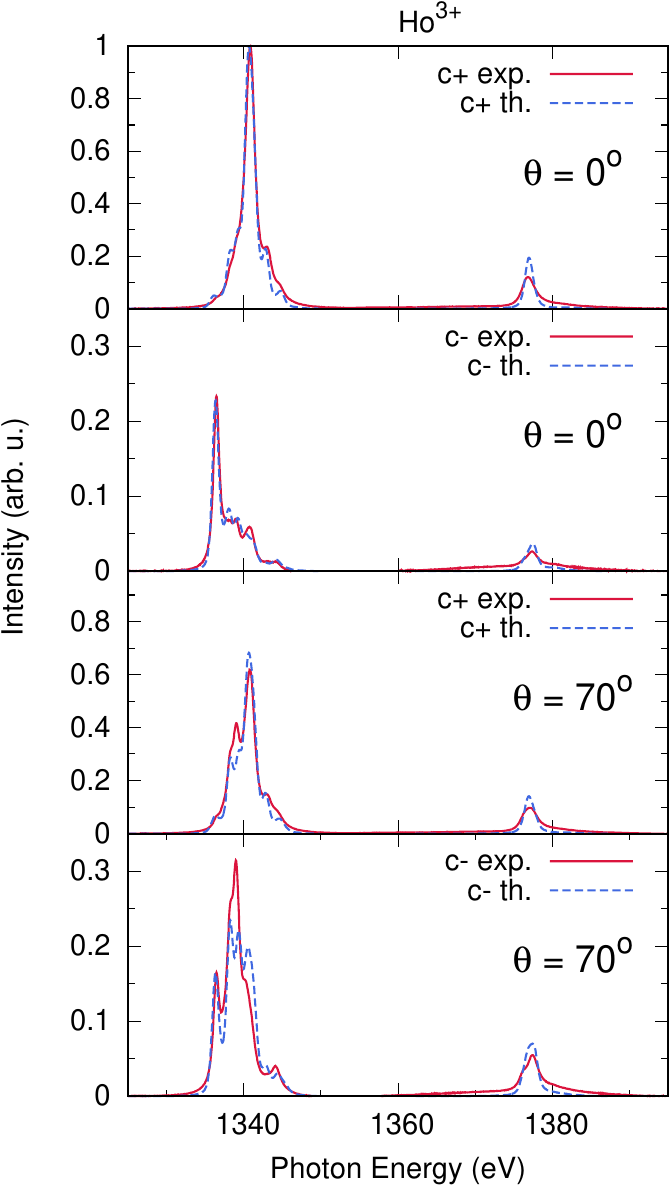}
\caption{\label{fig:hoau2_xmcd_spectra}
Set of four X-ray absorption spectra (both circular light polarizations and both
external magnetic field directions). Solid and dashed curves represent,
respectively, the experimental data and the theoretical simulation
assuming Ho$^{2+}$ (left) and Ho$^{3+}$ (right) valence states.}
\end{figure}

The possibility of a Ho$^{2+}$ valence state can be dismissed, as the basic features of the X-ray
spectra it yields differ from the observed ones (see
Fig.~\ref{fig:hoau2_xmcd_spectra} left-hand panels) \cite{bib:singha17}.
Therefore, in the following we focus the analysis on Ho$^{3+}$.
Since the $D_{6h}$ symmetry of the HoAu$_2$ monolayer is broken by the moir\'e pattern
formed on the substrate, the Ho atoms are subject to a $C_{3v}$ CF that varies spatially
with the local stacking. Being X-ray absorption spectroscopy a space-averaged technique,
the fitted parameters are to be interpreted as those of an effective 3-fold symmetric CF.
In particular, a trigonal symmetry implies that there are crossed CF terms, namely those
corresponding to the $B_{43}$ parameter. These terms provide the Eq.~\ref{eq:Hsia}
eigenstates with a mixed $\ket{J_z}$ character that eventually makes
magnetic moments unstable under quantum transitions \cite{bib:donati14}.
In a initial run, the Wybourne $B_{20}, B_{40}$, and $B_{43}$ parameters
are fit according to the procedure described in section \ref{sec:methods_multiplet}.
An ampliation of the search space to the higher-order
parameters $B_{30}, B_{60}, B_{63}$  does not improve the quality
of the fit and is thus not considered in the analysis.
The best-fit CF parameters are listed in Table~\ref{tab:hoau2_CFfit}
(see also Table~\ref{tab:hoau2_slater} for further parameters of the hamiltonian).
The right-hand panels of Fig.~\ref{fig:hoau2_xmcd_spectra} show that there is
good agreement between the experimental and simulated spectra.

Diagonalization of the spin Hamiltonian for this CF at a small zero external
field (0.01\,T) shows that the ground state $\Psi_0$ and the lowest-energy excited states ${\Psi_i} (i>0)$
of Ho$^{3+}$ belong to the $J=8$ multiplet (states with $J \neq 8$ contribute $<5$\%).
Their quantum numbers and decomposition into $\ket{J_z}$ states are summarized in
Table~\ref{tab:hoau2_diag} and the eigenenergies are shown in
Fig. 1(c) of the main paper.

$\Psi_0$ is a doublet with a configuration lead by $\pm 8$ states and $\langle J_z \rangle \simeq \pm 6$,
in agreement with previous sum rules analyses~\cite{Fernandez20Nanoscale}.
The first excited state ${\Psi_1}$ at 4.9\,meV is another doublet lead by a $\pm 7$ state.
As the 3-fold symmetric CF terms mixes terms $\ket{J_z \pm 3}$,
the projection  $\langle J_z \rangle$
cannot be quenched by quantum tunneling transitions,
as it is the case of Ho/Cu(111) \cite{bib:donati14}.
Such quenching scenario occurs for ${\Psi_2}$, which lies much higher in energy (38.7\,meV).
This electronic structure conferes the HoAu$_2$ a strong out-of-plane anisotropy
and magnetic bistability to individual Ho$^{3+}-4f$ momenta.
Note that the $\langle J_z \rangle$ quenching, due to a $\ket{\pm 6}$ state, happens in Ho/Cu(111)
at 2.4\,meV in Ho/Cu(111) \cite{bib:donati14}
and even at the GS of Ho/Pt(111) \cite{bib:donati14,bib:karbowiak16,bib:steinbrecher16}.
The main difference with respect to the adatom geometry is that the
two-dimensional alloy structure conferes
a more intense crystal field to Ho$^{3+}$ (see  Table~\ref{tab:hoau2_CFfit}).
In particular, the anisotropy-related parameter $B_{20}$ is larger
by one order of magnitude  and $B_{43}$ by two orders.


\begin{table}[!ht]
\caption{\label{tab:hoau2_CFfit}
Stevens crystal field parameters fitted for HoAu$_2$/Au(111) spectra at 6\,T
(values in the Wybourne convention are given in parenthesis),
compared to literature values for Ho adatoms on Cu(111) and Pt(111).
The last row indicates the ground state character.}
\begin{ruledtabular}
\begin{tabular}{ccccc}
          &  HoAu$_2$/Au(111) & Ho/Cu(111)\footnotemark[1]  & Ho/Pt(111)\footnotemark[1]   \\
\hline
$B_{0}^2$ ($B_{20}$) &  -1.92\,meV (1.73\,eV)     &  -0.056\,meV   & -0.14\,meV        \\
$B_{0}^4$ ($B_{40}$) &  2.5\,$\mu$eV (-0.305\,eV) &   0            &  1.06\,$\mu$eV   \\
$B_{3}^4$ ($B_{43}$) &  -0.125\,meV (-2.11\,eV)   &  5.9\,$\mu$eV  &  0                \\
GS $\ket{J_z}$       & $\ket{\pm 8}$              &  $\ket{\pm 8}$ & $\ket{\pm 6}$     \\
\end{tabular}
\end{ruledtabular}
\footnotetext[1]{From Ref.~\cite{bib:donati14}.}
\end{table}

\begin{table}[!ht]
\caption{\label{tab:hoau2_diag}
Eigenergies, average quantum numbers and configuration in terms of the
leading $\ket{J_z}$ state contributions (only weights
larger than 0.1 are shown) of the Ho$^{3+}-4f$ lowest energy states
($J=8$ multiplet) for the optimized crystal field in HoAu$_2$.}
\begin{ruledtabular}
\begin{tabular}{ccccc}
              & $\Psi_0^\pm$ \footnotemark[1] & $\Psi_1^\pm$ \footnotemark[2] & $\Psi_2$ \footnotemark[3] & $\Psi_3$ \footnotemark[4]  \\
\hline
Energy (meV)  & 0   & 4.91  & 38.7   &  55.7   \\
$L_z$         & $\pm 4.29$   & $\pm 3.52$   &  0   &  0   \\
$S_z$         & $\pm 1.45$   & $\pm 1.16$   &  0   &  0   \\
$T_z$         & $\pm 0.064$  & $\pm 0.045$  &  0   &  0   \\
\end{tabular}
\end{ruledtabular}
\footnotetext[1]{$|\braket{\Psi_0^+|8}|^2 = |\braket{\Psi_0^-|-8}|^2 = 0.59 $;
                 $|\braket{\Psi_0^+|5}|^2 = |\braket{\Psi_0^-|-5}|^2 = 0.27 $ }
\footnotetext[2]{$|\braket{\Psi_1^+|7}|^2 = |\braket{\Psi_1^-|-7}|^2 = 0.56 $;
                 $|\braket{\Psi_1^+|4}|^2 = |\braket{\Psi_1^-|-4}|^2 = 0.31 $ }
\footnotetext[3]{$|\braket{\Psi_2|6}|^2 = |\braket{\Psi_2|-6}|^2 = 0.27$;
                 $|\braket{\Psi_2|6}|^2 = |\braket{\Psi_2|-6}|^2 = 0.20$ }
\footnotetext[4]{$|\braket{\Psi_3|6}|^2 = |\braket{\Psi_3|-6}|^2 = 0.31$;
                 $|\braket{\Psi_3|6}|^2 = |\braket{\Psi_3|-6}|^2 = 0.18$ }
\end{table}


\section{Magnetization curves of the pristine  $\mathrm{HoAu}_2$ and $\mathrm{GdAu}_2$ substrates}
\label{sec:magcurve_model_subs}

\begin{figure}
\includegraphics[width=0.9\columnwidth,angle=0]{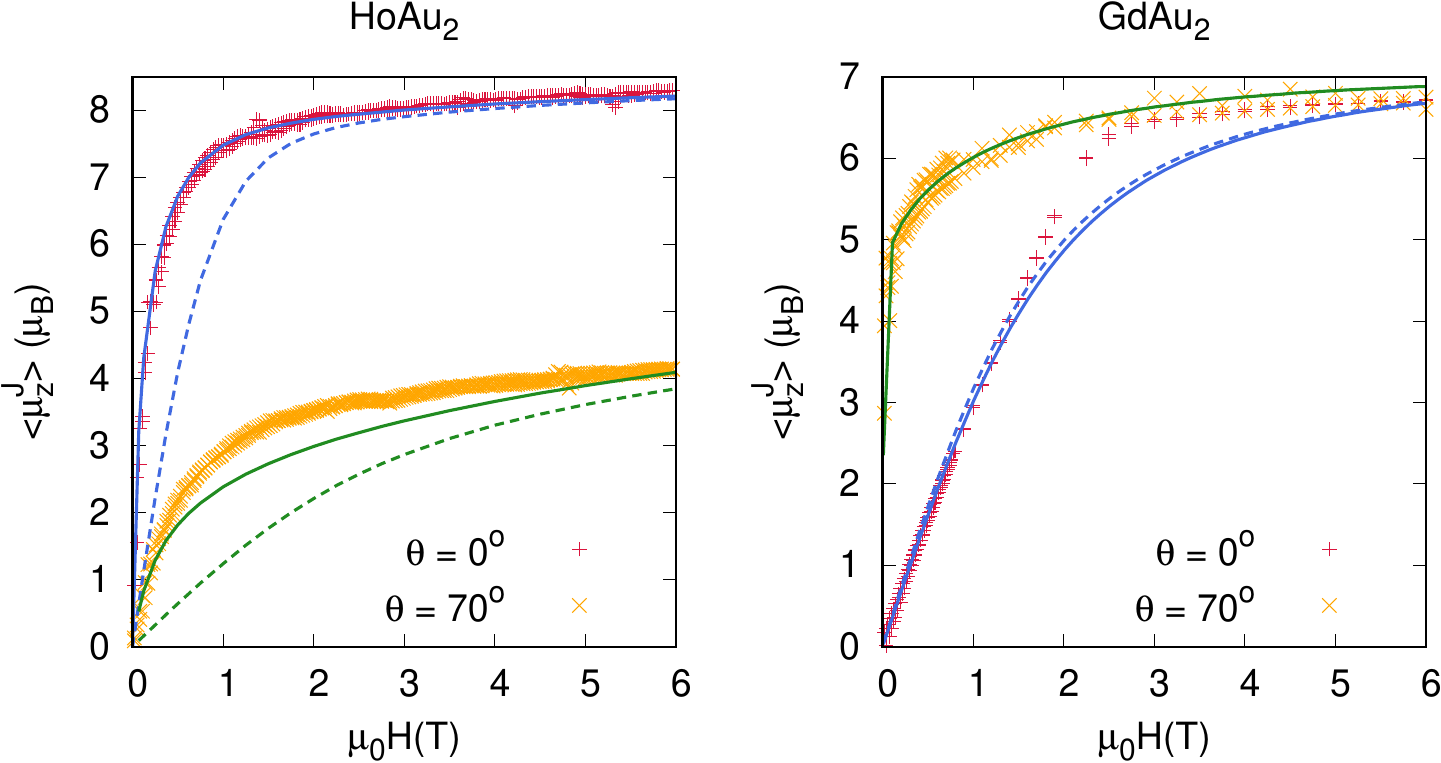}
\caption{\label{fig:magcurve_subs}
Experimental magnetization curves (symbols), measured as the
$M_5$ peak intensity in the XMCD spectrum as a function of the applied magnetic fields.
The models (solid lines, blue for normal and green for grazing incidence),
described in section \ref{sec:magcurve_model_subs}
for each system, use effective fields (Eq.~\ref{eq:Beff_subs}) with Weiss parameters
$\gamma_\perp = 0.095$ and $\gamma_\parallel = 0.75$\,T$/\mu_B$ for Ho,
$\gamma_\perp = -0.02$ and $\gamma_\parallel = 0.38$\,T$/\mu_B$ for Gd.
The dashed lines show, as a reference, the result of using
$\gamma_\perp = \gamma_\parallel = 0$.}
\end{figure}

As described in the previous section, the $4f$ orbital of Ho$^{3+}$ conferes a
pronounced perpendicular single-ion anisotropy (SIA) character to the behaviour of HoAu$_2$.
The HoAu$_2$ magnetization curve is modelled with Eq.~\ref{eq:Jzave_magcurve}
adding an extra Zeeman contribution to each eigenergy in the form of a self-consistent
Weiss mean-field, which accounts for the RKKY exchange interactions
between the Ho$^{3+}$ total momenta $\mathbf{J}$  mediated by itinerant $d$-electrons.
The effective field acting on them is
\begin{equation}
\mathbf{B}_{eff} = \mu_0\mathbf{H} + \hat\gamma \cdot \langle \vec\mu_J \rangle
\label{eq:Beff_subs}
\end{equation}
where the tensor $\hat\gamma$ is fitted to the $M_5$ peak intensity as a function of the applied
field $\mu_0\mathbf{H}$, which can be considered as the experimental magnetization curves.
Due to the exchange interaction anisotropy within the HoAu$_2$ monolayer,
the $\hat\gamma$ tensor components are different for moments aligned parallel and perpendicular to the surface
($\gamma_\parallel$ and $\gamma_\perp$, respectively).
These parameters are fitted to match overall the experimental curves, normalized to
the $\mu_{J,z}$ values (the subindex $z$ indicates the projection along the applied field direction)
extracted from the XAS and XMCD sum rules in Ref.~\cite{Fernandez20Nanoscale},
8.3 and 4.14\,$\mu_B$ for perpendicular and grazing incidences, respectively.
$\gamma_\perp$ is extracted from the normal incidence data and kept fixed in the
fit of the grazing incidence data, where $\gamma_\parallel$ is obtained.
The results are shown in Fig.~\ref{fig:magcurve_subs}.
Although the $d$-band effect on the exchange is at play, too, the SIA effect is
the dominant one, as reflected by the $\hat \gamma=0$ result, shown by the
dashed lines in Fig.~\ref{fig:magcurve_subs}.

For the GdAu$_2$ case we only consider the exchange anisotropy
in the modelling. Here the magnetization curves are well described simply by the Brillouin
function for $B_{eff}$ fields (Eq.~\ref{eq:Beff_subs}) and $J=7/2$,
albeit with significant exchange anisotropy:
an intense ferromagnetic Weiss field is needed to fit the grazing incidence curve,
while its contribution at normal incidence is almost negligible (see Fig.~\ref{fig:magcurve_subs}).
Note that the linear regime in the experimental $\theta=0^\circ$ curve persists until 2.5\,T, where
saturation is almost complete. This feature, not captured by our model,
might be due to magnetic domain effects.

\section{Anisotropy of the bare  $\mathrm{CuPc}$ molecule}
\label{sec:cupc_anis}

The planar symmetry of the singly-occupied $d_{x^2-y^2}$ orbital of CuPc protects its spin
$S=1/2$~\cite{Rosa_1994,Liao01} upon adsorption of monolayers or submonolayers
of CuPc on several metallic surfaces.
Despite the $S=1/2$ character, both CuPc MLs and thin layers display
a pronounced out-of-plane anisotropy, which is
originated from the strong out-of-plane anisotropy of its orbital momentum
and spin dipole operator projections on the field direction, $L_z$ and
$T_z$, respectively. This results in a significant attenuation of the spin
effective moment $\mu^{S}_{eff} = 2S + 7T$ at grazing
incidence~\cite{Stepanow_PRB2010_CuPc_Ag,Gargiani_PRB,Bartolome}.

\begin{figure}
\includegraphics[width=0.9\columnwidth,angle=0]{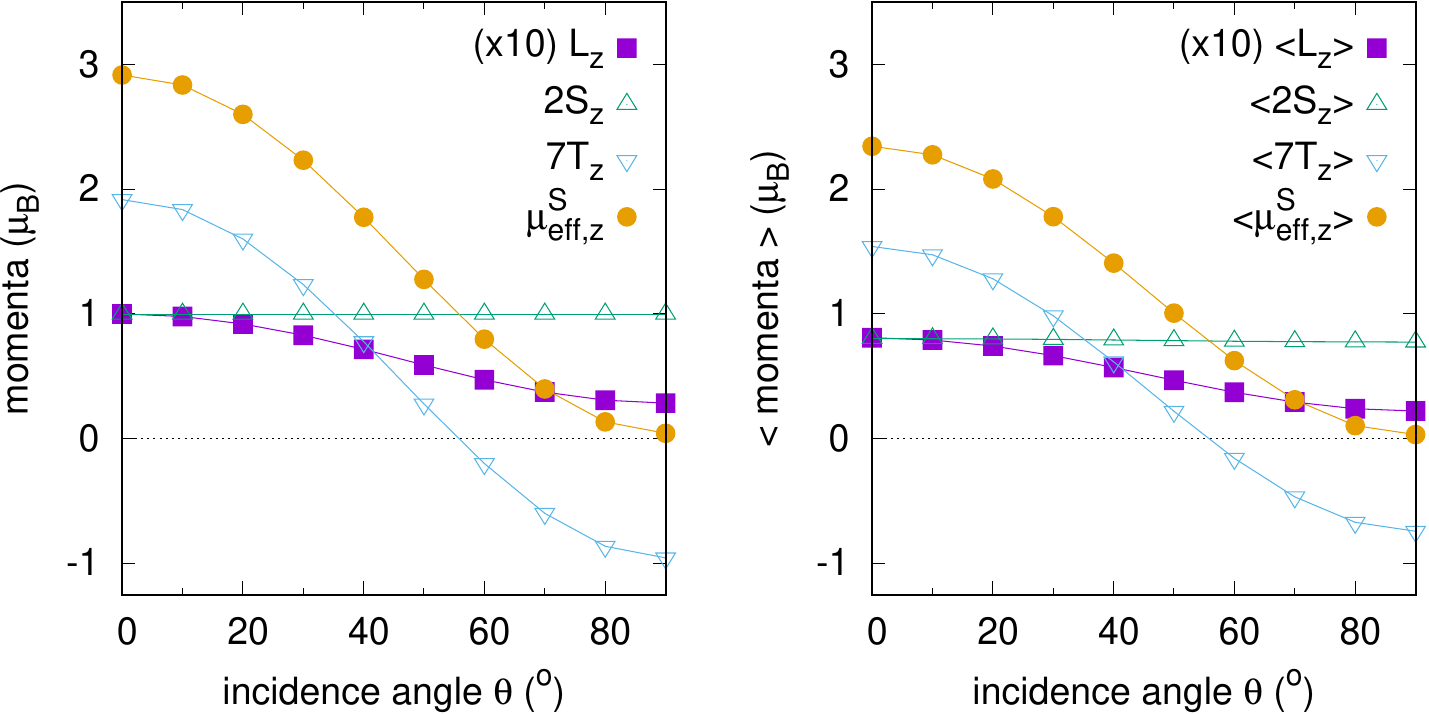}
\caption{\label{fig:CuPc_Tz}
Projected $L_z$ (values scaled by a factor 10), $2S_z$, $7T_z$ and $\mu^{S,z}_{eff} = 2S_z + 7T_z$ momenta on the applied field direction for the
bare CuPc molecule, using the orbital energies of Ref.~\cite{Rosa_1994}.
Left: $T=0$\,K saturation values. Right: expectation values at $\mu_0 H=6$\,T and $T=4$\,K.}
\end{figure}

The crystal field for the Cu$^{2+}$ ion in CuPc is established by
the $d$-orbital energy levels. For the free molecule, we take these levels
from calculated optical transitions in Ref.~\cite{Rosa_1994}:
$\epsilon (d_{x^2-y^2}) - \epsilon (d_{z^2}) = 3.48$\,eV,
$\epsilon (d_{z^2}) - \epsilon (e_g) = 0.15$\,eV, and
$\epsilon (e_g) - \epsilon (d_{xy}) = 0.40$\,eV.
By exact diagonalization of the Cu$^{2+}$ single-ion Hamiltonian with a CF given by
these parameters, we find that the ground state with $S=1/2$ has non-negligible $L_z$ and,
most important, a significant $T_z$ at normal incidence.
The variation of these quantum numbers with the polar angle of incidence is shown
in Fig.~\ref{fig:CuPc_Tz}, both the $T=0$\,K saturation values and the expected ones
at a 6\,T field with $T=4$\,K.
The bare molecule $T_z$ and $\mu^{S,z}_{eff}$ values are larger than when adsorbed on
Ag(001)~\cite{Stepanow_PRB2010_CuPc_Ag}, HoAu$_2$ and GdAu$_2$
(this work, Table I) which suggests that, although the CuPc anisotropy is reduced,
the out-of-plane easy axis of magnetization is preserved
upon adsorption in planar physisorbed configurations.


\section{Magnetic moments of $\mathrm{CuPc}$ adsorbed on $\mathrm{HoAu}_2$ and $\mathrm{GdAu}_2$ }
\label{}

Magnetic moments $\mu^{L}$ and $\mu^S_{eff}$  of CuPc adsorbed on HoAu$_2$ and GdAu$_2$, and the magnetic moment ratios between the OOP
(0$^{\circ}$) and the IP (70$^{\circ}$) values. The values are  obtained from the sum
rules analysis of the XMCD measurements performed at 6\,T. Additionaly, the corresponding values for the systems CuPc/Ag(100) and CuPc/Au(110) are included for comparison. Simulated bare CuPc values are also included in the table.

\begin{table*}[tb!]
\caption{\label{tab:sumrules} In this table the magnetic moments $\mu_{L}$ and $\mu_S{}_{\textnormal{eff}}$ of
CuPc/GdAu$_2$ and CuPc/HoAu$_2$ are compared with a multilayer of CuPc (5 ML), theory calculations and other data extracted from the literature for CuPc ML on coinage substrates.The values of CuPc/GdAu$_2$ and CuPc/HoAu$_2$ are obtained from the sum
rules analysis of the XMCD measurements performed at 6\,T. The error range is 
of 0.005 $\mu_{B}$ for $\mu_{L}$(0$^{\circ}$) and $\mu_{L}$(70$^{\circ}$). For $\mu_S^\textnormal{eff}$(0$^{\circ}$) the error range amounts 
to 0.05 $\mu_{B}$ and for $\mu_S^\textnormal{eff}$(70$^{\circ}$) to 0.02 $\mu_{B}$.  
}
    \begin{tabular}{c|c|c|c|c|c|c}
                      & $\mu_{L}$(0$^{\circ}$) &
$\mu_S{}_\textnormal{eff}$(0$^{\circ}$) & $\mu_{L}$(70$^{\circ}$) &
$\mu_S{}_\textnormal{eff}$(70$^{\circ}$) &
$\mu_{L}$(0$^{\circ}$)/$\mu_{L}$(70$^{\circ}$) &
$\mu_S{}_\textnormal{eff}$(0$^{\circ}$)/$\mu_S{}_\textnormal{eff}$(70$^{\circ}$)\\
      \hline
          CuPc/HoAu$_2$                                & 0.1 & 1.36 &
0.04 & 0.17 & 2.5 & 8.1 \\
          CuPc/GdAu$_2$                                & 0.067& 1.05 &
0.06  & 0.28 & 1.1 & 3.8  \\
          5 ML CuPc                                    & 0.12 & 1.16 &
0.1  & 0.38 & 1.3 & 3.1  \\
          CuPc/Au(110)~\cite{Gargiani_PRB}             & 0.09 & 0.87 &
0.02  & 0.11 & 4.6 & 8.0  \\
          CuPc/Ag(100)~\cite{Stepanow_PRB2010_CuPc_Ag} & 0.10 & 1.67 &
0.03  & 0.33 & 3.1 & 5.0  \\
          CuPc (theory)                                & 0.08 & 2.34 &
0.03  & 0.31 & 2.8 & 7.6  \\

    \end{tabular}
\end{table*}



\newpage

\bibliography{magnetCuPc}